\documentclass[12pt]{iopart}
\usepackage{graphicx}
\usepackage{cite}
\usepackage{xcolor}

\usepackage{hyperref}

\begin{document}
\title[Retrieval of the nuclear motion...]{Retrieval of the nuclear motion in a molecule from photoelectron momentum distributions using non-Born-Oppenheimer quantum dynamics and deep learning}

\author{N I Shvetsov-Shilovski$^{1,2}$ and M Lein$^{2}$}

\address{$^{1}$Rudolf Bembenneck Gesamtschule Burgdorf, Vor dem Celler Tor 50, 31303 Burgdorf, Germany}

\address{$^{2}$Institut f\"{u}r Theoretische Physik, Leibniz Universit\"{a}t Hannover, 30167 Hannover, Germany}

\ead{n79@narod.ru}

\begin{abstract}
By using a neural network that takes momentum distributions of photoelectrons produced in strong-field ionization as input, we retrieve the time-dependent bond length of a dissociating one-dimensional H$_{2}^{+}$ molecule. The photoelectron momentum distributions are calculated from the direct numerical solution of the non-Born-Oppenheimer time-dependent Schr\"{o}dinger equation. We simulate two setups: first, molecules prepared in the first excited electronic state, second, a pump-probe scheme starting from the ground state. We show that in both schemes a neural network trained on momentum distributions calculated for frozen nuclei retrieves the time-dependent bond length with an absolute error of $0.2-0.4$~a.u. We find that a neural network, when applied to photoelectron momentum distributions obtained within the pump-probe scheme, can be used for the retrieval of the equilibrium internuclear distance and ground-state population. This opens new perspectives for extracting electronic properties of molecules from electron momentum distributions using deep learning.

\end{abstract}

\noindent{\it Keywords\/}: time-dependent bond-length, deep learning, strong-field ionization, photoelectron momentum distributions\\

\submitto{\jpb}
\maketitle
% \ioptwocol

\section{Introduction}

The past few years have witnessed remarkable progress in the development of artificial intelligence and its subfield machine learning (ML). ML is currently widely used in all branches of physics: from astrophysics to soft-matter, medical, and biological physics, see, e.g., Refs.~\cite{Carleo2019,Suresh2024} for reviews. This is due to the fact that ML is capable of processing huge amounts of data and drawing conclusions based on these data. Furthermore, ML methods are used in order to find complex mappings that depend on many variables and are hidden in the input data. 

ML has also begun to be used in strong-field physics, which studies the interaction of strong laser radiation with atoms and molecules. The interaction of quantum systems with intense laser pulses results in a variety of highly non-linear phenomena, including above-threshold ionization (ATI), high-order harmonic generation (HHG), and sequential and nonsequential double ionization, see, e.g., Refs.~\cite{BeckerRev2002,MilosevicRev2003,FaisalRev2005,FariaRev2011,Graefe2016,Lin2018} for reviews. Among applications of ML in strong-field physics are the prediction of the flux of high-order harmonics \cite{Gherman2018}, reconstruction of the intensity and the carrier-envelope phase (CEP) of short laser pulses from 2D images (frequency-resolved optical gating traces \cite{Zahavy2018} and dispersion scan traces \cite{Kleinert2019}), the efficient implementation of the trajectory-based Coulomb-corrected strong-field approximation (TCSFA) \cite{Yang2020} (the TCSFA method is developed in Refs.~\cite{Yan2010,Yan2012}), the retrieval of the femtosecond pulse parameters based on photoelectron momentum distributions (PMDs) produced in the ATI process \cite{Szoldra2023}, and the simulation of the macroscopic HHG driven by structured laser beams \cite{Hernandez2023}. Another remarkable example is the prediction of HHG spectra for model diatomic and triatomic molecules for randomly chosen laser intensity, internuclear distance, and orientation of the molecule using ML, as well as the solution of the inverse problem: retrieval of the molecular and laser parameters and determination of the type of molecule based on the high-harmonic spectrum or time-dependent dipole acceleration \cite{Lytova2023}. Recently, ML methods have been applied to the identification of the geometrical structure of molecules in laser-induced electron diffraction (LIED) \cite{Liu2021}, creating movies of attosecond charge migration using high harmonic spectroscopy \cite{He2022},  and the determination of the internuclear distance in a molecule from a given PMD \cite{Shvetsov2022,Shvetsov2023,Shvetsov2024}. These latter applications can be used as tools for time-resolved molecular imaging.  

Time-resolved molecular imaging aims to visualize electronic and molecular dynamics in real time. The development of such techniques is extremely important for many branches of science and technology, including chemistry, biology and material science, as they provide a better understanding of various processes in molecules. Among the well-known methods of time-resolved molecular imaging are optical pump-probe spectroscopy, time-resolved x-ray absorption spectroscopy, time-resolved electron diffraction and time resolved x-ray diffraction, time-resolved scanning tunneling spectroscopy and others (see Ref.~\cite{Agostini2016} for a review). Some of the processes studied in strong-field physics are applicable to molecular imaging. For example, the LIED (see, e.g., Refs.~\cite{Meckel2008,Blaga2012,Pullen2015}) and strong-field photoelectron holography (SFPH) methods \cite{Huismans2011} analyze electron momentum distributions produced in the ATI process.

These methods are based on the fact that PMDs encode a lot of time-resolved information about the ion and the photoelectron. Two different kinds of electron wave packets generated in strong-field ionization contribute to the PMDs. The low-energy part of the distributions is formed to a large extent by direct electrons that do not return to their parent ions. In contrast to this, the high-energy part of PMDs is created by rescattered electrons that are driven back by the oscillating laser field to their parent ions and scatter from them by angles close to $180^{\circ}$. The LIED method retrieves static and time-resolved information from the high-energy part of PMD, i.e., from the rescattered electrons. In contrast to this, SFPH analyzes the holographic patterns emerging due to the interference of direct and rescattered electrons in the low-energy part of PMDs.    

Both the LIED and SFPH techniques have been intensively developed over the last decade, see Ref.~\cite{Giovannini2023} for a recent review of the LIED method and Refs.~\cite{Walt2017,Keil2017,Maurer2018,He2018a,He2018b,Tan2019,Brennecke2019} for the modern achievements in studies of SFPH. In Refs.~\cite{Shvetsov2022,Shvetsov2023,Shvetsov2024} a convolutional neural network (CNN) was used for the retrieval of the static and time-dependent internuclear distance in a two-dimensional (2D) H$_2^{+}$ molecule based on the whole PMD (low and high-energy parts). It was shown that the CNN can retrieve the fixed internuclear distance with an absolute error of less than $0.1$~a.u. \cite{Shvetsov2022}.  In contrast to the method of Ref.~\cite{Liu2021}, where a CNN was trained on the two-dimensional (2D) differential cross sections, the information encoded in the electron momentum distributions was exploited directly in Ref.~\cite{Shvetsov2022}. The corresponding PMDs were obtained from the numerical solution of the time-dependent Schr\"{o}dinger equation (TDSE), and therefore a number of approximations that are usually made in the LIED and SFPH techniques were avoided.  By using the transfer learning technique \cite{Goodfellow2016}, the CNN \cite{Shvetsov2022} has been made applicable to distributions calculated for parameters it was not explicitly trained for, including focal-volume averaged PMDs and PMDs from another molecular species \cite{Shvetsov2023}. 

In Ref.~\cite{Shvetsov2024} a CNN has been applied for reconstruction of the time-dependent bond length in the dissociating 2D H$_2^{+}$ molecule based on electron momentum distributions obtained within a pump-probe scheme. As in Refs.~\cite{Shvetsov2022,Shvetsov2023}, the PMDs were obtained from the solution of the TDSE, and the motion of the nuclei was treated classically, semiclassically and quantum mechanically. In all these cases, the neural network trained on PMDs calculated for frozen nuclei predicts the time-dependent internuclear distance with an absolute error of $0.2$-$0.3$~a.u. However, the quantum approach of Ref.~\cite{Shvetsov2024} is approximate: it does not account for the motion of the nuclei during the probe laser pulse. It is therefore of interest to investigate whether ML is able to reconstruct the time-dependent bond length from PMDs obtained by treating the dynamics of the molecule fully quantum mechanically without any approximations.  In this paper we address the formulated problem by employing non-Born-Oppenheimer simulations. For simplicity, we consider a dissociating one-dimensional (1D) H$_2^{+}$ molecule. We use two different schemes. First, we assume an instantaneous excitation of the molecule, i.e. the initial state is a nuclear wave packet prepared in the first excited electronic state. Second, we consider a pump-probe scheme starting from the molecular ground state, i.e., the molecule is excited by a pump pulse and ionized by a time-delayed probe pulse. We demonstrate that neural networks trained on PMDs calculated with moving nuclei in both schemes retrieve the internuclear distance with good accuracy. We then study the application of a neural network trained on momentum distributions calculated for frozen nuclei to the retrieval of the time-dependent internuclear distance. We find that in both instantaneous excitation and pump–probe schemes, a neural network trained on PMDs for fixed nuclei can retrieve the time-dependent internuclear distance with an absolute error of about $0.2$–$0.4$~a.u. Finally, we demonstrate that deep learning can be used for the reconstruction of not only nuclear motion but also the electronic properties of the molecule. In the pump–probe scheme, we use a neural network trained on PMDs for fixed nuclei to reconstruct the electronic ground-state population at the end of the pump pulse.

The paper is organized as follows. In Sec.~2 we discuss the method used to solve the TDSE and to calculate the PMDs as well as the architecture of neural networks used in the present study. In Sec.~3 we train and apply our neural networks to both schemes, and we analyze the obtained results. The conclusions are given in Sec.~4. Atomic units are used throughout the paper unless indicated otherwise. 
 
\section{Methods}

The time-dependent Schr\"{o}dinger equation for a model H$_2^{+}$ molecule, in which the electronic and nuclear motions are restricted to one spatial dimension, is given by
\begin{eqnarray}
i\frac{\partial}{\partial t}\Psi\left(x,R,t\right)&=\left\{-\frac{1}{2\mu}\frac{\partial}{\partial x^{2}}-\frac{1}{m_{p}}\frac{\partial^2}{\partial R^2}-i\frac{m_{p}+1}{m_{p}}A\left(t\right)\frac{\partial}{\partial x}+V_{pe}\left(x,R\right)\right.\nonumber\\
&\left.+\frac{1}{\sqrt{R^2+\alpha_{1}}}\right\}\Psi\left(x,R,t\right),
\label{tdse_main}
\end{eqnarray}
where $x$ is the electron coordinate, $R$ is the internuclear distance, $\Psi\left(x,R,t\right)$ is the wave function, $m_{p}$ is the proton mass, $A\left(t\right)$ is the vector potential of the laser pulse, $\mu$ is the reduced mass for the electron motion, $\alpha_{1}$ is the soft-core parameter for the proton-proton interaction, and the potential $V_{pe}$ describes the interaction between the electron and both protons. It is worth noting that for a homonuclear diatomic molecule, the laser field does not couple to the $R$ coordinate, see Ref.~\cite{Hiskes1961}. The velocity gauge is used in Eq.~(\ref{tdse_main}). The potential $V_{pe}$ is chosen as 
\begin{equation}
V_{pe}\left(x,R\right)=-\frac{1}{\sqrt{\left(x-\frac{1}{2}R\right)^2+\alpha_{2}}}-\frac{1}{\sqrt{\left(x+\frac{1}{2}R\right)^2+\alpha_{2}}},
\label{potential_pe}
\end{equation}
where $\alpha_{2}$ is the soft-core parameter. Following Ref.~\cite{Kulander1996}, we have chosen $\alpha_{1}=0.03$ and $\alpha_{2}=1.0$. 

The vector potential $A\left(t\right)$ of the ionizing laser pulse is given by
\begin{equation}
A_{\mathrm{probe}}\left(t\right)=\left(-1\right)\frac{F_0}{\omega}f\left(t\right)\cos \left[\omega\left(t-t_{c}\right)\right],
\label{vecpot1}
\end{equation}
where $\omega$ is the laser frequency, $F_{0}$ is the field-strength amplitude, $t=t_{c}$ is the center of the laser pulse, $\tau_{L}$ is the pulse duration, and the envelope $f\left(t\right)$ is given by
\begin{equation}
f\left(t\right)=\cases{\sin^2\left[\frac{\omega\left(t-t_{c}+\frac{\tau_{L}}{2}\right)}{2n_{p}} \right]&for $t_{c}-\frac{\tau_{L}}{2}\leq t \leq t_{c}+\frac{\tau_{L}}{2}$\\
0&otherwise.}
\label{envelope1}
\end{equation}
Here, $n_p$ is the number of the optical cycles within the pulse and, therefore, $\tau_{L}=\left(2\pi/ \omega\right)\cdot n_{p}$. We perform our simulations for $\varphi=0$.

Equation (\ref{tdse_main}) or similar equations has been used in many studies of H$_{2}^{+}$ interacting with strong laser pulses, see, e.g., Refs.~\cite{Kulander1996,Picon2011,Mosert2015}. In order to solve the TDSE (\ref{tdse_main}), we use the Feit-Fleck-Steiger split-operator method \cite{Feit1982}. The computational box extends over $x\in\left[-400,400\right]$~a.u. and $R\in\left[0,80\right]$~a.u. The grids in $x$ and $R$ directions consist of 4096 and 2048 points, respectively, which corresponds to the grid spacings $\Delta x=0.1954$~a.u. and $\Delta R=0.0391$~a.u. The TDSE is propagated in time with the time step $\Delta t=0.0184$~a.u. We use absorbing boundaries in the $x$ direction, in order to prevent unphysical reflections of the wave function from the boundary of the computational grid, i.e., at every time step we multiply the wave function by the mask:
\begin{equation}
\label{cases}
M\left(x\right)=\cases{1&for $|x|\leq x_b$\\
\exp\left[-\beta\left(x-x_b\right)^2\right]&for $|x| > x_b$\\},
\label{mask}
\end{equation}
where $\beta=10^{-4}$ and $x_b=300$~a.u.
The potential $V_{pe}$ is neglected for $|x|>x_{b}$, but the interaction between the protons described by the term $1/\sqrt{R^2+\alpha_{1}}$ term in Eq.~(\ref{tdse_main}), is taken into account in this region. Indeed, large values of $x$ do not necessarily imply large values of $R$. Due to the large extent of the computational box in the $R$ direction, no absorbing mask is applied in this direction. The joint momentum distributions for electron and nuclei are obtained by using the mask method, see Refs.~\cite{Lein2002,Tong2006}. The electron momentum distributions are calculated by integrating the joint momentum distributions over the proton momenta. The ground-state wave function of the model H$_{2}^{+}$ molecular ion is obtained by imaginary time propagation, providing the ground-state energy $E_{0}=-0.78$~a.u., which is consistent with Refs.~\cite{Picon2011,Mosert2015}. 

Similar to Ref.~\cite{Shvetsov2024}, our first approach involves instantaneous excitation of the H$_{2}^{+}$ molecule. More specifically, the time propagation of the TDSE (\ref{tdse_main}) starts with the wave function $\Psi_{0}\left(x,R\right)=\chi_{0}\left(R\right)\phi_{1}\left(x,R\right)$. Here $\phi_{1}\left(x,R\right)$ is the first excited electronic state for fixed $R$, and $\chi_{0}\left(R\right)$ is the nuclear ground-state wave function for the electronic ground state. Accordingly, the wave function $\phi_{1}\left(x,R\right)$ is a solution of the following time-independent Schr\"{o}dinger equation
\begin{equation}
\left[-\frac{1}{2\mu}\frac{\partial^{2}}{\partial x^{2}}+V_{pe}+\frac{1}{\sqrt{R^2+\alpha_1}}\right]\phi_{n}\left(x,R\right)=\varepsilon_{n}\left(R\right)\phi_{n}\left(x,R\right),
\label{tdse_stat}
\end{equation}
where the internuclear distance $R$ is a parameter. In practice, we use imaginary time propagation for the time-dependent analogue of (\ref{tdse_stat}) to find $\phi_{1}\left(x,R\right)$. To enforce the correct symmetry, we symmetrize $\phi_{0}\left(x;R\right)$ [$1s\sigma_{g}$ state] and antisymmetrize the $\phi_{1}\left(x,R\right)$ [$2p\sigma_{\mu}$ state] at every step of the imaginary time propagation.  The ground state energy $\varepsilon_{0}\left(R\right)$ is required to formulate the Schr\"{o}dinger equation for the nuclear motion:
\begin{equation}
\left[-\frac{1}{m_{p}}\frac{\partial^{2}}{\partial R^{2}}+\varepsilon_{0}\left(R\right)\right]\chi_{0}\left(R\right)=\varepsilon_{0}\chi_{0}\left(R\right).
\label{tdse_nuc}
\end{equation}
Ionization of the H$_2^{+}$ molecule is performed by a strong and short probe pulse acting after a certain time delay $t_{c}$, see Eq.~(\ref{vecpot1}), that determines the internuclear distance at the ionization time.  We first solve the TDSE (\ref{tdse_main}) without the probe pulse $\left[A\left(t\right)=0\right]$ and calculate the average internuclear distance $\langle R \rangle$ as a function of time:
\begin{equation}
\langle R \rangle = \int dx \int dR R |\Psi\left(x,R,t_{C}\right)|^{2}.
\label{r_aver}
\end{equation}
Knowing this function allows us to assign an internuclear distance as a label to a PMD calculated for a given delay of the probe pulse.  Then, the TDSE (\ref{tdse_main}) is solved in the presence of the probe pulse, i.e. $A\left(t\right)=A_{\mathrm{probe}}\left(t\right)$ to obtain the PMDs. To obtain the training data from fixed-nuclei calculations, we solve the 1D TDSE
\begin{eqnarray}
&i\frac{\partial}{\partial t}\Psi\left(x,t\right)=\left\{-\frac{1}{2}\frac{\partial^2}{\partial x^2}-iA\left(t\right)\frac{\partial}{\partial x}-\frac{1}{\sqrt{\left(x-\frac{1}{2}R\right)^2+\alpha_2}}\right.\nonumber\\
&\left.-\frac{1}{\sqrt{\left(x+\frac{1}{2}R\right)^2+\alpha_2}}\right\}\Psi\left(x,t\right)
\label{tdse_1d}
\end{eqnarray}
for ionization from the first excited state at different values of the internuclear distance $R$. 

As a second approach, we consider the well-known pump-probe scheme. More specifically, we solve the TDSE (\ref{tdse_main}) with the vector potential given by $A\left(t\right)=A_{\mathrm{pump}}\left(t\right)+A_{\mathrm{probe}}\left(t\right)$,
\begin{equation} 
A_{\mathrm{pump}}\left(t\right)=\frac{F_{01}}{\omega_1}f_{1}\left(t\right)\sin\left(\omega_{1}t\right)
\label{vec_pot_pump}
\end{equation}
where $F_{01}$ and $\omega_{1}$ are the field strength and the frequency of the pump pulse, respectively, and 
\begin{equation}
f_{1}\left(t\right)=\cases{\sin^2\left(\frac{\omega_{1}t}{2n_{p1}}\right)&for $0\leq t \leq \tau_{pump}$\\
0&otherwise.}.
\label{envelope2}
\end{equation}
Here, $n_{p1}$ and $\tau_{\mathrm{pump}}=\left(2\pi/ \omega_{1}\right)\cdot n_{p1}$ denote the number of cycles and the duration of the pump pulse, respectively. Since the envelope (\ref{envelope2}) reaches its maximum at $t=\pi n_{p1}/ \omega_{1}$, the delay between the pump and probe pulses is given by $t_{c}-\pi n_{p1}/\omega_{1}$.

The PMDs obtained from the solution of the TDSE (with moving nuclei or fixed nuclei, see below) are used for training of neural networks. More specifically, we split the set of calculated PMDs into the training and test sets using the ratio $0.75:0.25$. It should be noted that the resulting PMDs are functions of only one variable - the electron momentum component along the polarization direction. Therefore, application of a CNN is not the only option, and a fully connected neural network (FCCN) can also be applied to retrieve the bond length from PMDs. In the following, we shall use both the FCCN and CNN and compare the respective results with each other. We consider the PMDs in the range $-3.0<k_{x}<3.0$~a.u., and we normalize the distributions to their maximum values. As a result, every PMD is represented by a set of $765$ values between $0$ and $1$. These sets are given to a neural network as input. In the present study, we employ two different FCNNs, see Figs.~1~(a) and (b). Apart from an input layer, the first FCCN consists of only two fully connected layers with $765$ and $1512$ neurons, respectively, a dropout layer preventing overfitting (see, e.g., Ref.~\cite{TraskBook}), and a last fully connected layer that predicts the output of the network, see Fig.~1~(a). The number of neurons in this layer corresponds to the number of output quantities predicted by the network. The first two fully connected layers apply the rectified linear unit (ReLU) activation function: RELU$\left(x\right)=\textrm{max}\left(0,x\right)$. The architecture of the second FCCN is shown in Fig.~1~(b). While the first two fully connected layers of this network are similar to those of the first FCCN, it additionally includes a third layer with 1512 neurons, a fourth layer with 1512 neurons, and a fifth layer with 512 neurons. Consequently, this network contains a larger number of trainable parameters. We use the MATLAB package \cite{Matlab} to implement the neural network. We split the training data into minibatches consisting of a certain number of distributions in order to use one minibatch for each iteration of the training process. The goal of training is to minimize the loss function (in our case the mean squared error), which characterizes the deviation between predictions of the neural network and the known labels of the training PMDs, i.e., internuclear distances. The training of the network starts using the learning rate $l_r=5\cdot10^{-3}$, which is decreased by a factor of $10$ after $20$ training epochs. About $60$ epochs are needed for convergence of the loss function. The training data are shuffled before each epoch, and therefore each PMD leads to an unbiased modification of the neural network. For training, we use the stochastic gradient descent (SGD) optimizer. 

\begin{figure}[h]
\centering
\includegraphics[width=.98\textwidth]{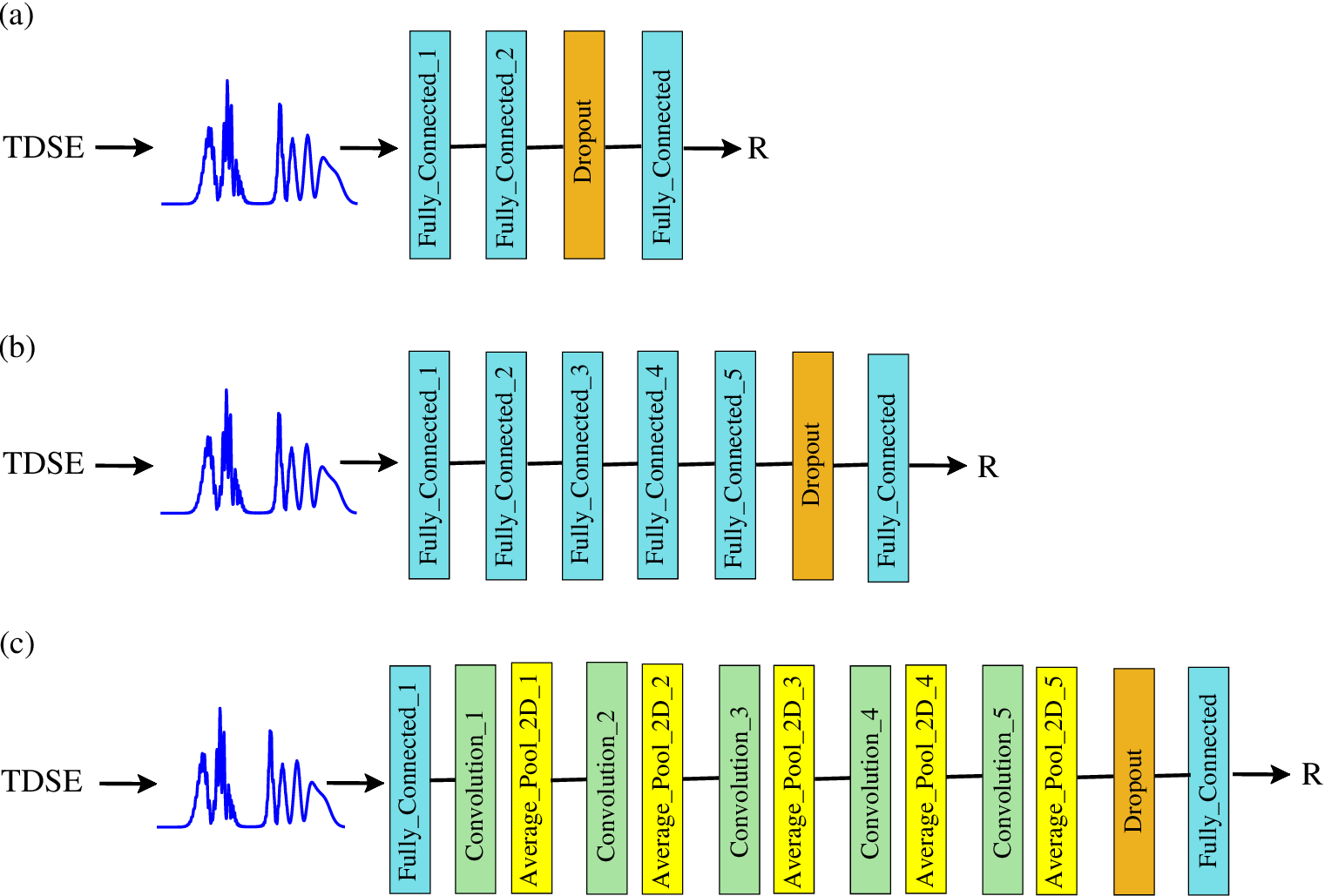}
\caption{The architecture of the (a), (b) FCCNs and (c) CNN used for the retrieval of the bond length $R$. Both neural networks take 1D electron momentum distributions normalized to their maximum value as input; see text.}
\label{arch}
\end{figure}

Similar to the networks used in studies \cite{Shvetsov2022,Shvetsov2023,Shvetsov2024}, our present CNN consists of five pairs of convolutional layers and average pooling layers, dropout layer, and a fully connected player calculating the output of the network, see Fig.~1~(b). In contrast to CNNs~\cite{Shvetsov2022,Shvetsov2023,Shvetsov2024}, the pairs of convolutional and average pooling layers are preceded by a fully connected layer with $\left(2\cdot765\right)=1530$ neurons. Both convolutional and average pooling layers work in only one dimension. Each convolutional layer consists of 32 filters, with each filter having a size of 2 points.  Apart from the convolution of the input, the convolutional layers apply the ReLU activation function. The size of the pooling region in average pooling layers is also equal to 2 points. We find that for the problem at hand, good performance of the CNN with this architecture is achieved for a minibatch consisting of 30 images and the initial learning rate equal to $10^{-2}$. The learning rate is again decreased by a factor of ten after 20 epochs.

\section{Results and discussion}

\subsection{Retrieval of the time-dependent bond length for instantaneous excitation}

For simplicity, we start the application of deep learning with the case of fixed laser intensity. We solve the TDSE (\ref{tdse_main}) starting with instantaneous ionization and using $N=310$ random delay times of the probe pulse and an intensity of $2.52\times10^{14}$ W/cm$^2$. The range of delay times corresponds to $3.61<R<12.00$~a.u. The resulting PMDs [see, e.g., Fig.~2~(a)]  were used to train a FCCN aimed at the retrieval of the internuclear distance at the ionization time. The loss function converges after about $40$ training epochs, and the neural network recognizes the internuclear distance with a mean absolute error (MAE) of $0.07$~a.u.

\begin{figure}[h]
\centering
\includegraphics[width=.80\textwidth]{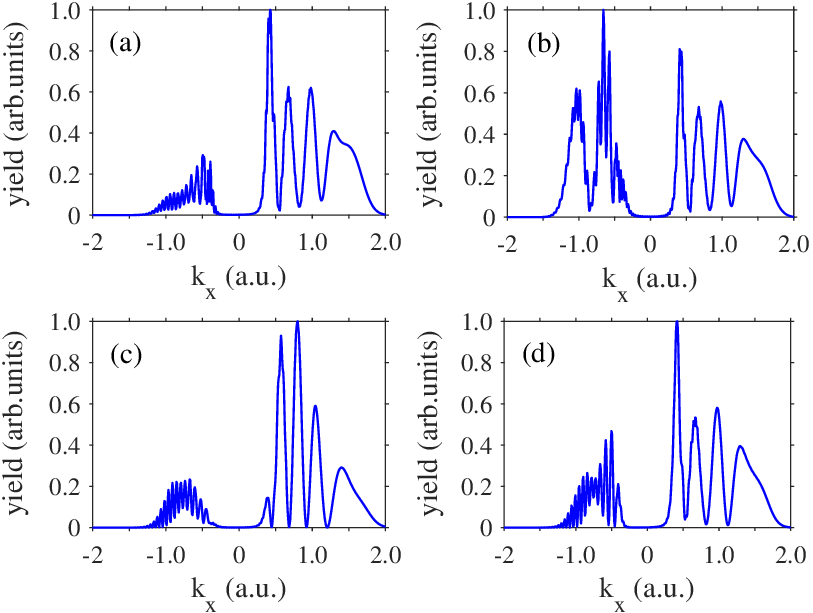}
\caption{Electron momentum distributions for ionization of the 1D H$_2^{+}$ molecule calculated from the direct numerical solution of the TDSE for the internuclear distance $R=9.26$~a.u. 
The molecule is ionized by a laser pulse with duration $n_{p}=2$ cycles, peak intensity $2.5\times 10^{14}$ W/cm$^2$, and wavelength $800$~nm. (a) Momentum distribution obtained from the moving-nuclei TDSE for instantaneous excitation of the molecule and ionization at $\langle R \rangle=9.26$~a.u. (b) Distribution calculated within the pump-probe scheme. The wavelength of the pump pulse is $148$~nm ($\omega=0.368$~a.u.), the peak laser intensity is $8.4\times10^{12}$~W/cm$^2$, and duration of this pulse is $n_p=3$ optical cycles. (c) Distribution obtained from the solution of the fixed-nuclei TDSE (\ref{tdse_1d}) for ionization of the ground state. (d) Distribution calculated from Eq.~(\ref{tdse_1d}) for ionization of the first excited state. The distributions are normalized to their maximum value.}
\label{ints}
\end{figure}

We then allow the intensity to vary in the range $\left[1.0, 4.0\right]\times10^{14}$ W/cm$^2$, and we calculate $N=2000$ PMDs for random internuclear distances and laser intensities. We train the FCNN shown in Fig.~1~(a). The minibatches consist of $100$ distributions. The neural network predicts the internuclear distance and the laser intensity with MAEs equal to $0.08$~a.u. and $0.05$ W/cm$^2$, respectively [see Figs.~3~(a) and 3~(b)]. For comparison, in addition to the FCNN, we also train a CNN for this purpose. The CNN predicts $R$ and $I$ with the MAEs equal to $0.094$~a.u. and $0.068$ W/cm$^2$, respectively. Therefore, the use of the CNN does not provide an advantage over a FCCN for the problem at hand. 

\begin{figure}[h]
\centering
\includegraphics[width=.80\textwidth]{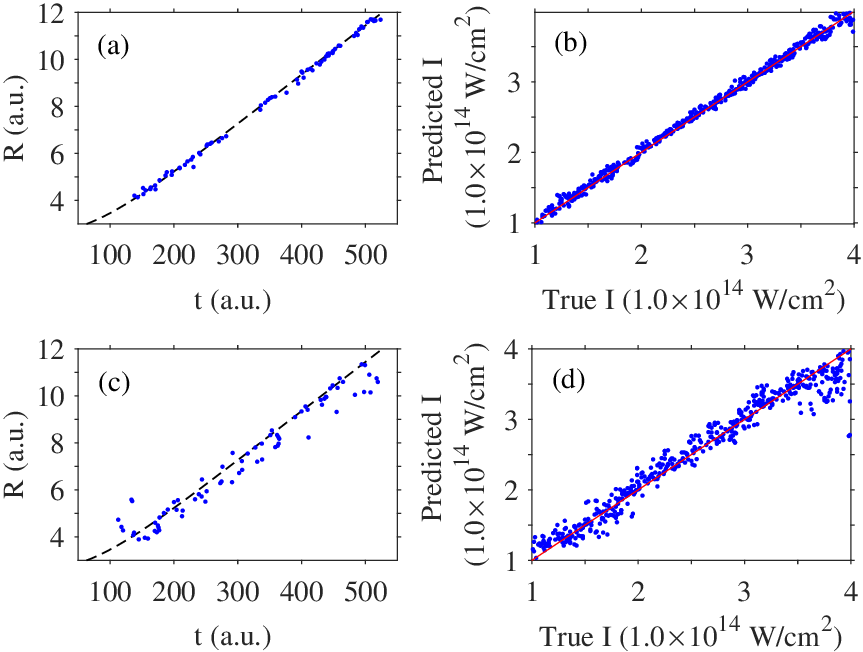}
\caption{Plots of predicted vs true internuclear distances [(a),(c)] and laser intensities [(b),(d)] illustrating the performance of neural networks for the case of instantaneous excitation. (a) and (b) correspond to an FCCN trained on a set of distributions calculated from the TDSE~(\ref{tdse_main}) with moving nuclei. (c) and (d) show the performance of an FCCN trained on a set of PMDs with fixed nuclei, i.e., obtained from the solution of Eq.~(\ref{tdse_1d}), for ionization from the first excited state. The dashed curves in (a) and (c) show the time-dependent expectation value of the bond length.}. 
\label{ints}
\end{figure}

As a next step, neural networks trained on distributions calculated for fixed internuclear distances are applied to PMDs obtained for moving nuclei. A set of $N=2000$ fixed-nuclei PMDs for ionization from the first excited state is used to train an FCCN [see Fig.~1~(b)] to retrieve $R$ and $I$. An example of such a distribution is shown in Fig.~2~(d). This distribution should be compared with the PMD for ionization from the ground state shown in Fig.~2~(c). The neural network shows good performance on a test set of PMDs with fixed nuclei: the corresponding MAEs for $R$ and $I$ are equal to $0.10$~a.u. and $0.04$~W/cm$^2$, respectively. However, its performance on distributions obtained for moving nuclei is considerably worse, although it remains acceptable: MAE$\left(R\right)=0.48$~a.u., MAE$\left(I\right)=0.12$~W/cm$^2$, see Figs.~3~(c) and 3~(d). A CNN trained on the distributions for fixed nuclei shows similar performance. These results are easy to understand, considering that fixed nuclei PMDs differ from momentum distributions calculated for moving nuclei, see Figs.~2~(a) and 2~(c). 

\subsection{Retrieval of the time-dependent bond length in the pump-probe scheme}

We perform the calculations for the pump pulse frequency $\omega=0.3068$~a.u., which corresponds to the resonance between the $\sigma_g$ and $\sigma_{\mu}$ states in our model molecule, the field strength $F_{01}=0.0155$~a.u. (intensity of $8.4\times10^{12}$~W/cm$^2$), and $n_{p_{1}}=3$. As in Sec.~3.1, we first calculate the time-dependent average internuclear distance $\langle R \rangle$  as a function of time in the absence of the probe pulse. This expectation value is calculated after projecting out the electronic-vibrational ground state at the end of the pump pulse \cite{Picon2011}. We next solve the TDSE (\ref{tdse_main}) with the vector potential (\ref{vec_pot_pump}) and we calculate a set of $N=2000$ momentum distributions corresponding to various values of the time delay $\tau$, and therefore various values of the internuclear distances $\langle R \rangle$, lying in the range $\left[3.98, 11.96\right]$~a.u. An example of a PMD is shown in Fig.~2~(b). Every PMD is labelled with an internuclear distance $R$ in accordance with the function $\langle R \rangle\left(t\right)$. This set of PMDs is used for training of the FCCN shown in Fig.~1~(b) and a CNN aimed at the recognition of the internuclear distance and the laser intensity. We find that the FCCN retrieves the internuclear distance with an MAE of 0.13~a.u. and the laser intensity with an MAE of $0.05\times10^{14}$~W/cm$^2$. The CNN shows similar performance: $\textrm{MAE}\left(R\right)=0.15$~a.u. and $\textrm{MAE}\left(I\right)=0.044\times10^{14}$~W/cm$^2$.

At the next stage of our study, we apply neural networks trained on the distributions calculated for fixed nuclei to PMDs obtained in the pump-probe scheme. The FCCN shows poor performance in recognizing the internuclear distance when applied to PMDs obtained in the pump-probe scheme. The corresponding MAE for the internuclear distance is equal to $1.13$~a.u., and the laser intensity can be reconstructed with an MAE of $0.25\times10^{14}~$W/cm$^2$. The CNN trained on PMDs with fixed nuclei demonstrates similar performance. These results are substantially worse compared to the case of instantaneous excitation (see Sec.~3.1.). In order to simplify the task of the neural network, we fix the laser intensity to $2.52\times10^{14}$~W/cm$^2$. We calculate a set of corresponding fixed nuclei PMDs, and we train an FCCN using this set. However, this fixed-intensity neural network performs only marginally better than the previous ones. For PMDs calculated for moving nuclei, it predicts the internuclear distance with an MAE of $1.03$~a.u. 

These results could have been expected. Indeed, PMDs calculated from the TDSE (\ref{tdse_main}) in the pump-probe scheme for given values of $\langle R \rangle$  at the ionization time differ from the PMDs obtained from Eq.~(\ref{tdse_1d}) for the same internuclear distances. This applies to ionization of both the ground and excited states in the TDSE (\ref{tdse_1d}) for fixed nuclei [cf. Fig.~2~(b) with Figs.~2~(c) and 2~(d)]. Analysis of the populations in the case of the TDSE (\ref{tdse_main}) shows that the population of the ground electronic-vibrational state at the end of the pump pulse is equal to $0.92$. This means that the pump pulse creates a superposition of the ground state (with a larger weight) and the first excited state (with a much smaller weight). This suggests that a set of PMDs defined as
\begin{equation}
\textrm{PMD}_{i}=w_{i}\textrm{PMD}_{\phi_{0,i}}\left(R_{0,i}\right)+\left(1-w_{i}\right)\textrm{PMD}_{\phi_{1,i}}\left(R_{1,i}\right)
\label{set_pmd}
\end{equation}
can be used for training of a fixed-nuclei neural network aimed at the retrieval of the time-varying bond length. Here $i$ is the number of a distribution in the set, $\textrm{PMD}_{\phi_{0,i}}\left(R_{0,i}\right)$ and $\textrm{PMD}_{\phi_{1,i}}\left(R_{1,i}\right)$ are electron momentum distributions calculated from Eq.~(\ref{tdse_1d}) for ionization of the ground and first excited states, respectively, and $w_{i}$ is the weight of $\textrm{PMD}_{\phi_{0,i}}$. Note that the interatomic distances $R_{0,i}$ and $R_{1,i}$, for which the distributions $\textrm{PMD}_{0,i}$ and $\textrm{PMD}_{1,i}$ are calculated, do not necessarily have to be the same. Therefore, a set of PMDs (\ref{set_pmd}) is determined by the choice of $R_{0,i}$, $R_{1,i}$, and $w_{i}$.

We first fix $R_0$ to the equilibrium distance equal to $2.59$~a.u. (for $\alpha_2=1.0$), and we calculate a set of $\textrm{PMD}_{\phi1}\left(R_{1,i}\right)$ for $R_{1,i}$ distributed randomly in the range $4.1\leq R_{1,i}\leq 11.5$~a.u. The idea behind this choice is that only the wave packet in the excited electronic state undergoes dissociation. For simplicity, we again fix the laser intensity to $2.52\cdot10^{14}$~W/cm$^2$. The choice of the weights $w_{i}$ is of crucial importance for the training procedure, and it turns out to be a challenging task. This is easy to understand if we compare the maximum of the $\textrm{PMD}_{\phi_{0}}(2.59~\textrm{a.u})$ equal to $5.93\cdot10^{-10}$ to the maxima of $\textrm{PMD}_{\phi_1}\left(R_1\right)$ for different values of $R_1$. These latter maxima only weakly depend on $R_1$. For example, $\max\left[\textrm{PMD}_{\phi_1}\left(R_1\right)\right]$ for $R_1=4.0$~a.u., $R_1=6.0$~a.u., and $R_1=8.0$~a.u. are equal to $2.96\cdot10^{-6}$, $2.89\cdot10^{-6}$, and $2.29\cdot10^{-6}$, respectively. The choice of the range for the weights $w_{i}$ is determined by the requirement that both terms in the sum (\ref{set_pmd}) have comparable values. Simple estimates, complemented by training of a few neural networks with different trial sets of PMDs, shows that the best results can be achieved for $w_{i}\in\left[0.99,0.99999\right]$, i.e., for $\left(1-w_{i}\right)\in\left[10^{-5},0.01\right]$. This suggests applying the neural network to a set of PMDs calculated for a lower intensity of the pump pulse than the intensity of $8.4\times10^{12}$ W/cm$^2$ used above. The pump-pulse intensity should be such that the ground-state population after the end of the pump pulse falls within the range of the weights used for training of the neural network. Therefore, we calculate another set of PMDs for the pump pulse intensity of   $3.29\times10^{10}$~W/cm$^2$ corresponding to the field strength $F_{01}=0.00097$~a.u. For this intensity, the ground-state population at the end of the pump pulse is $0.99948$.

As is commonly acknowledged, neural network training is inherently stochastic, see, e.g., Ref.~\cite{Goodfellow2016}. Different neural networks trained on the same training data set show differences in predicting internuclear distance. In the present case, this effect is particularly pronounced. For this reason, we train 10 different FCCNs on the same data set (\ref{set_pmd}), which consists of $5000$ distributions, and apply them to the test set of PMDs discussed above. The architecture of these neural networks is shown in Fig.~\ref{arch}~(b), and the minibatch size is set to 60 images. We find that these neural networks retrieve the time-dependent internuclear distance with an MAE of $0.48$~a.u. (averaged over neural networks). The performance of one selected FCCN is illustrated by Fig.~4~(a). Simultaneously, the FCCNs show slightly better performance on PMDs calculated for higher pump-probe intensity. For example, for the intensity of $8.4\times10^{12}$~W/cm$^2$, the average MAE is equal to $0.40$~a.u. This counterintuitive result along with the relatively low accuracy of the retrieved $R$ suggests that the approach should be modified.

\begin{figure}[h]
\centering
\includegraphics[width=.60\textwidth]{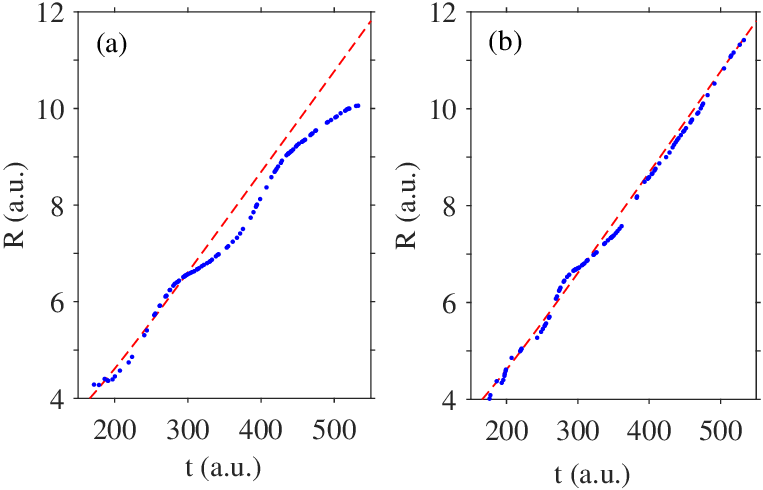}
\caption {Predictions of neural networks for the internucelar distance during dissociation (blue points) at different time delays compared to the time-dependent expectation value of the internuclear distance (red dashed curves) obtained from the solution of the TDSE (\ref{tdse_main}) for the pump-probe scheme. (a) Results from the FCCN trained on momentum distributions calculated from Eq.~(\ref{set_pmd}) for fixed $R_0=2.59$~a.u. The pump pulse intensity is $3.29\times10^{10}$ W/cm$^2$ . (b) Results from the FCCN trained on the set of distributions  (\ref{set_pmd}) with $R_{0,i}$ randomly distributed in the range $\left[1.0, 3.0\right]$~a.u. The PMDs of the test set are again calculated for the intensity of the pump pulse equal to  $3.29\times10^{10}$ W/cm$^2$.}
\label{ints}
\end{figure}

In our next approach, we vary not only $R_{1}$ but also $R_{0}$ in Eq.~(\ref{set_pmd}). More specifically, while the values $R_{1,i}$ vary again in the range $\left[4.1, 11.5\right]$~a.u., $R_{0,i}$ are randomly chosen from the range $\left[1.0, 3.0\right]$~a.u. It should be noted that $\textrm{max}\left[\textrm{PMD}_{\phi_0}\left(R_0\right)\right]$ changes significantly with increasing $R_0$. Indeed, $\max \textrm{PMD}_{\phi_0}\left(1.02\right)=1.14\cdot10^{-10}$, $\max \textrm{PMD}_{\phi_0}\left(2.02\right)=2.10\cdot10^{-10}$, and $\max \textrm{PMD}_{\phi_0}\left(3.02\right)=1.74\cdot10^{-9}$. This makes the problem of identifying a suitable interval for $w_{i}$ more difficult as compared to the previous case. Nevertheless, we find that the weights $w_{i}\in\left[0.999,  0.999999\right]$ provide the best results. For the pump-probe intensity of $3.29\times10^{10}$ W/cm$^2$, a set of $N=10$ FCNNs with the architecture shown in Fig~1~(b) trained on $10000$ PMDs (\ref{set_pmd}) with random weights in this interval predicts the time-dependent internuclear distance $R_{1,i}$ with an averaged MAE of $0.19$~a.u. Figure~4~(b) shows the performance of one FCNN selected from this set.

We also apply the same method for the retrieval of the internuclear distances $R_{0,i}$ and weights $w_{i}$ from pump-probe PMDs. In this case, the spread of predictions from different neural networks trained on the same data set is relatively small. The average internuclear distance $\langle R_{0,i}\rangle$ for the ground state delivered by a FCCN for the set of PMDs calculated for the pump pulse intensity of  $3.29\times10^{10}$ W/cm$^2$ is equal to $2.68$~a.u. This value is close to the equilibrium distance of $2.59$~a.u. in the Born-Oppenheimer approximation. Simultaneously, the neural network predicts the average weight  $\langle w_{i} \rangle$ of $0.99955$. This value is close to the ground-state population $0.99948$ at the end of the pump pulse. The obtained results show that the neural network trained on PMDs for frozen nuclei may be used for the retrieval of the bound-state populations and the internuclear distance in both dissociating and non-dissociating states. More research is needed to analyze the potential of this scheme in view of the fact that the ionization probability is strongly $R$-dependent even within the narrow range of the ground state nuclear distribution.

\section{Conclusions}
In conclusion, we have investigated the retrieval of the time-dependent internuclear distance of a dissociating 1D molecule using a deep neural network that takes PMDs produced by strong-field ionization as input. To this end, we either assumed an instantaneous excitation of the molecule to the first excited electronic state with subsequent dissociation or applied a pump-probe scheme. In the latter approach the nuclear motion is initiated by a pump pulse that excites the molecule. The dissociating molecule is ionized by the probe pulse coming after a certain time delay, which determines the internuclear distance at the time of ionization. In both cases the corresponding PMDs were calculated from direct numerical solution of the TDSE with non-Born-Oppenheimer treatment of the coupled electronic-nuclear motion. We have shown that a deep neural network (FCCN or CNN) trained on a few thousand PMDs retrieves the internuclear distance with an absolute error of about $0.1$~a.u. This applies to both FCCNs and CNNs. 

We have also applied a neural network trained on momentum distributions obtained at fixed internuclear distances. For instantaneous excitation of the molecule, the neural network retrieves the time-dependent internuclear distance with an MAE of $0.4$~a.u. For application in the pump-probe scheme, a network has been trained on incoherent sums of electron momentum distributions calculated for ionization from the electronic ground state and from the first excited state with frozen nuclei taken with randomly distributed weights. We have found that this neural network can reconstruct the time-dependent internuclear distance with an absolute error of $0.2$~a.u. This applies to both simulation schemes and both types of neural networks (FCCN and CNN) used in our study. Therefore, our results support the applicability of deep learning to the retrieval of the dissociation dynamics of molecules from electron momentum distributions.

Moreover, our work indicates that a neural network trained on PMDs with frozen nuclei may be used to reconstruct the populations of the electronic states. Therefore, this work demonstrates the application of deep learning to the retrieval of electronic properties in addition to nuclear motion. Advances in this direction are useful for the development of innovative tools for static and time-dependent molecular imaging.

\end{document}